\documentclass[]{spie}

\usepackage{amsmath,amsfonts,amssymb}
\usepackage{graphicx}
\usepackage{tabularx}
\usepackage[colorlinks=true, allcolors=blue]{hyperref}

\title{Automated Quantum Circuit Generation for Computing Inverse Hash Functions}

\author[a]{Elena R. Henderson}
\author[a]{Jessie M. Henderson}
\author[b]{William V. Oxford}
\author[a]{Mitchell A. Thornton}
\affil[a]{Darwin Deason Institute for Cyber Security, Southern Methodist University, \hspace{7em} 6425 Boaz Lane Dallas, TX 75205, USA}
\affil[b]{Anametric, Inc., Austin, TX 78652, USA}

\authorinfo{Send correspondence to ERH: erhenderson@smu.edu. Further author information: JMH: hendersonj@smu.edu, WVO: oxford@anametric.com, MAT: mitch@smu.edu}

\begin{document} 
\maketitle

\begin{abstract}
Several cryptographic systems depend upon the computational difficulty of reversing cryptographic hash functions.
Robust hash functions transform inputs to outputs in such a way that the inputs cannot be later retrieved in a reasonable amount of time even if the outputs and the function that created them are known.
Consequently, hash functions can be cryptographically secure, and they are employed in encryption, authentication, and other security methods.
It has been suggested that such cryptographically-secure hash functions will play a critical role in the era of post-quantum cryptography (PQC), as they do in conventional systems.
In this work, we introduce a procedure that leverages the principle of reversibility to generate circuits that invert hash functions.
We provide a proof-of-concept implementation and describe methods that allow for scaling the hash function inversion approach.
Specifically, we implement one manifestation of the algorithm as part of a more general automated quantum circuit synthesis, compilation, and optimization toolkit.
We illustrate production of reversible circuits for crypto-hash functions that inherently provide the inverse of the function, and we describe data structures that increase the scalability of the hash function inversion approach.
\end{abstract}


\section{Introduction}\label{sec:intro}
Hash functions transform inputs of one form to outputs of another via a mathematical function\cite{paar10}.
Ideally, every unique input is transformed to a unique output, and robust hash functions minimize collisions, which occur when two differing inputs map to the same output.
Since the 1970s, hundreds of hash functions have been used in applications such as digital signatures, password tables, key derivation, and blockchains\cite{preneel10,mironov05,zheng18,chaudhary19,wohlmacher00}.
Hash functions designed for such cryptographic applications must satisfy specific properties; for example, many hash functions are designed to be \textit{one-way} such that it is straightforward to move from input \textit{I} to output \textit{O}, but very difficult to recover \textit{I} when only \textit{O} and the hash function are known.
The process of moving from output to input is termed \textit{computation of an inverse hash function} or \textit{hash reversal}, and efficient techniques for reversing cryptographically-sound hash functions are widely researched.

We have developed a procedure for synthesizing reversible circuits that constitute hash function inverses.
The remainder of this article describes this method as follows:
We begin with background on cryptographic hashes, logical reversibility, and the toolkit that we use to automate production of inverse hash function circuits.
We then describe the procedure for generating the inverse hash circuits, and explain how they work.
Next, we present experimental results; as all of the functions we consider are small by cryptographic standards, we close by discussing avenues for further work, including additional applications and scalable function representations.

\section{Background}
\subsection{Cryptographic Hash Functions}
In this article, we limit our scope to hash functions defined for bitstrings.
Specifically, we consider hash functions that convert a fixed-length bitstring input to a fixed-length bitstring output.
There are several criteria that are useful for general-purpose hash functions; for example, it is generally desirable for differing inputs to map to differing outputs in as many cases as possible\cite{paar10}.
Requirements are more stringent for \textit{cryptographic hash functions}, which are those for which input bitstrings should remain unobtainable even when the corresponding output bitstrings and the function that generated them are known.

The study of cryptographic hash functions is both rich and complex, so a comprehensive survey is beyond the scope of this article\cite{preneel10,mironov05,sobti12}.
Here, it suffices to note that there are several properties that are oft-considered requirements for cryptographic hash functions\cite{preneel10,mironov05,sobti12}, including the following four.
First, a hash function is defined as \textit{one-way} if it is difficult to determine an input from an output when both the output and the function in question are known\cite{mironov05}.
Second, a hash function is defined as \textit{second-preimage resistant} if---when given one input---it is difficult to find a second input that maps to the same output\cite{mironov05}.
Third, a hash function is defined as \textit{collision resistant} if it is difficult to find two inputs that have the same output\cite{mironov05}.
Fourth and finally, a hash function is defined as satisfying the \textit{Avalanche Criteria} if it meets a two-part specification\cite{forrie90,mironov05}.
First, each output should differ from its corresponding input with a Hamming distance that is at least half of the number of output bits.
And second, for each pair of inputs that differs by a Hamming distance of one, the corresponding outputs should differ by a Hamming distance of at least half of the number of output bits.

In addition to properties specifying the relationship between inputs and outputs, cryptographic hash functions are generally understood to require hundreds of bits in inputs and outputs, because such bitstring lengths make brute force attacks impractical\cite{preneel10}.
But in this article, we will consider much smaller hash functions of at most 8-bits for two reasons.
First, we aim to present conceptual examples, which quickly become prohibitive as the number of bits in a hash function increases.
Second, the function specification format that we use for our experimental results (\texttt{.pla} format; see Section \ref{sub:preparing_the_function}) accommodates some functions with hundreds of inputs and outputs, but does not scale for functions with the properties of cryptographic hash functions.
Consequently, in Section \ref{sec:conclusion_and_future_work}, we discuss alternative function specification formats that would allow for representing large cryptographic hash functions.

\subsection{Logical Reversibility}\label{sub:logical_reversibility}
\textit{Logically-reversible circuits} are those for which each output has a unique input: knowing solely the outputs and the function that was applied to generate them allows for obtaining the inputs that led to those given outputs.
Unsurprisingly, logically-reversible circuits must be comprised of logically-reversible operations, which are perhaps best explained with an example.

\begin{figure} [ht]
   \begin{center}
   \begin{tabular}{c}
   \includegraphics[width=0.35\linewidth]{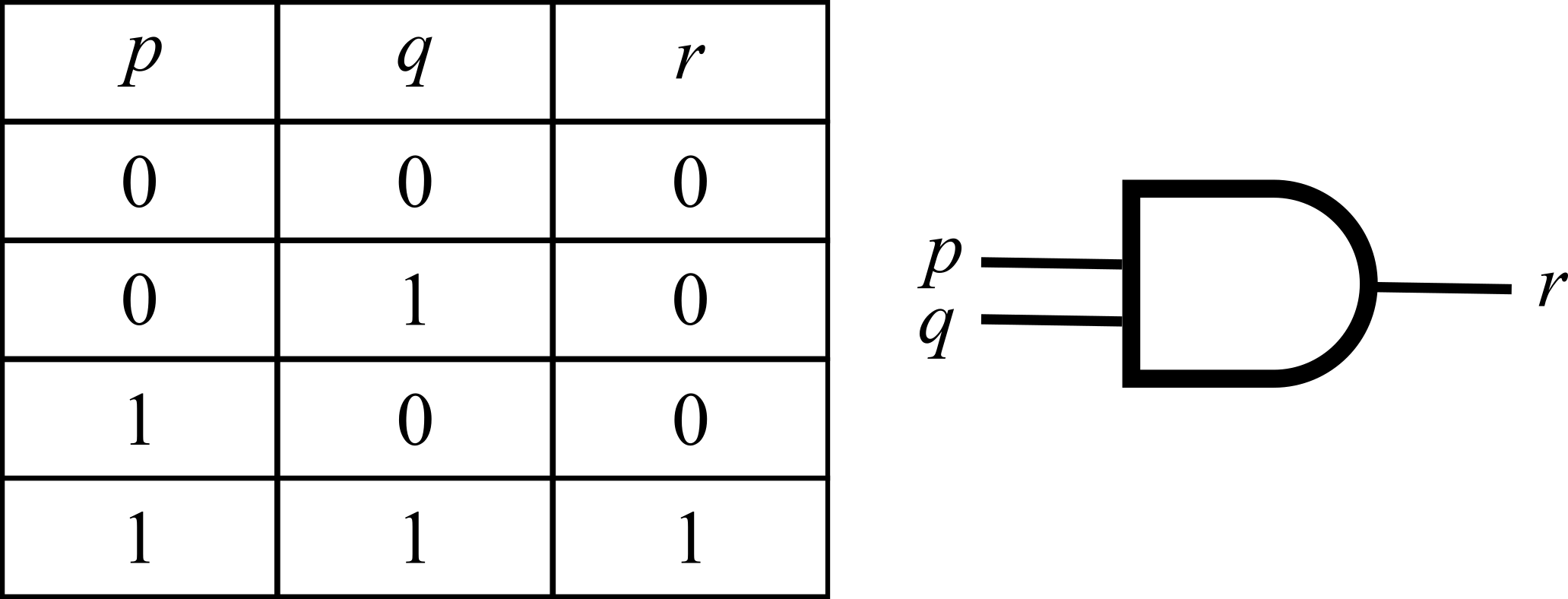}
   \end{tabular}
   \end{center}
   \caption{\label{fig:AND} 
Truth table and circuit representation for an irreversible AND operation.}
\end{figure}

\begin{figure} [ht]
   \begin{center}
   \begin{tabular}{c}
   \includegraphics[width=0.5\linewidth]{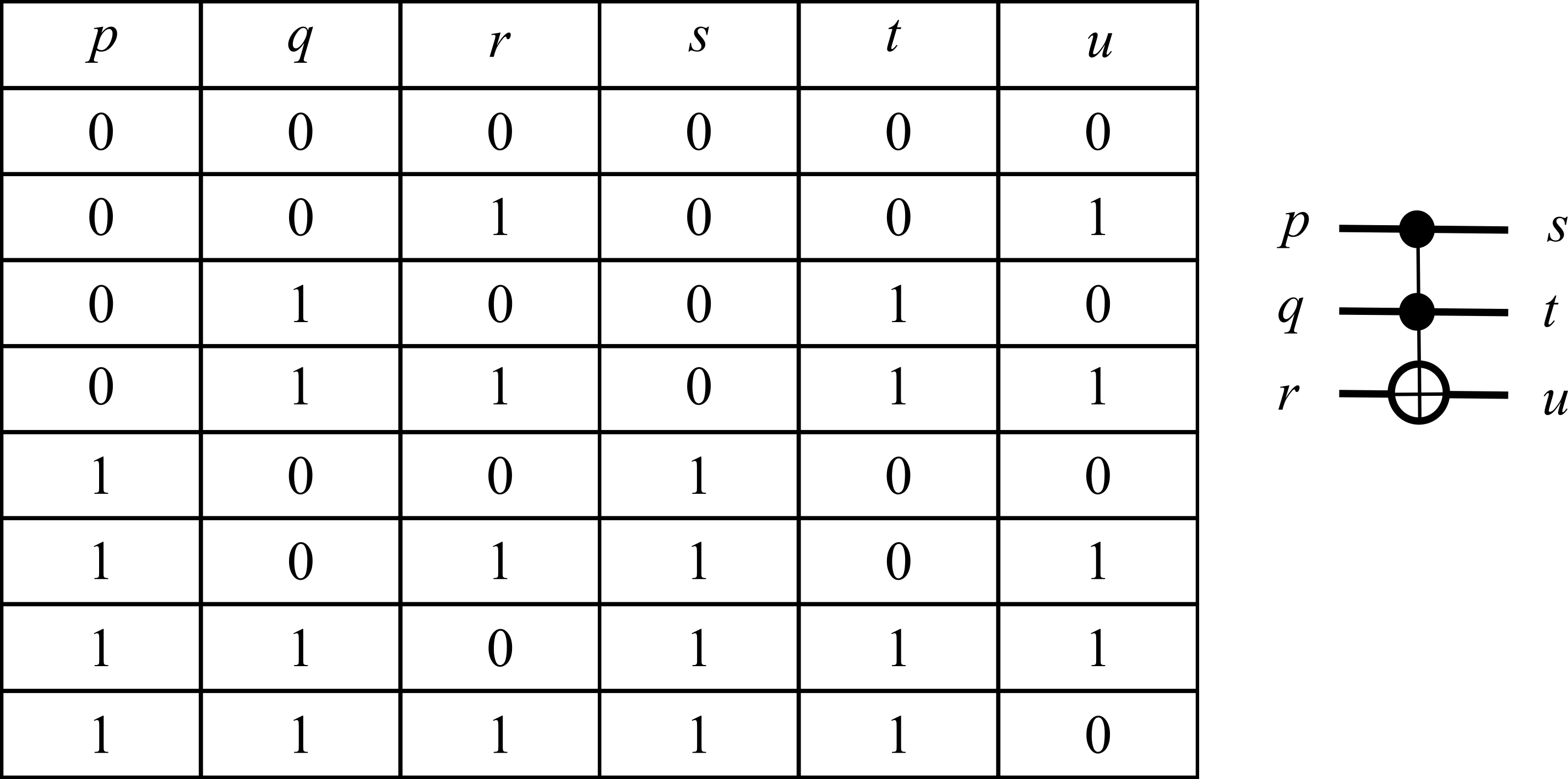}
   \end{tabular}
   \end{center}
   \caption{\label{fig:Toffoli} 
Truth table and circuit representation for a reversible Toffoli operation. The Toffoli operation is a controlled operation, in which a NOT is applied to $r$ if and only if both $p$ and $q$ are one. $p$ and $q$ are called the controls, and $r$ is termed the target of the operation. The values of controls do not change (meaning $s=p$ and $t=q$), but both values are included as outputs such that the operation is reversible.}
\end{figure}

Consider a two-input AND operation, which has the truth table and circuit diagram specified in Figure \ref{fig:AND}.
Given the output $r=0$ and the information that an AND operation was applied, one cannot know what the inputs were, since there are three possible input combinations associated with $r=0$: 1) $p=0$ and $q=0$; 2) $p=0$ and $q=1$; and 3) $p=1$ and $q=0$.
Consequently, the AND operation is not reversible.
Contrast this with the three-input Toffoli operation in Figure \ref{fig:Toffoli} and the outputs $s=1$, $t=1$, and $u=0$.
The only possible set of inputs that produces such outputs is $p=1$, $q=1$, and $r=1$.
Thus, the Toffoli operation is reversible, and we see that adding inputs and outputs can make an irreversible function reversible\cite{nielsen11}.

\subsection{Brief Introduction to \textit{MustangQ}}
Our inverse hash circuit-generation procedure is one application of a broader software design effort, so we briefly introduce \textit{MustangQ} before describing the reversal process.
\textit{MustangQ} is a synthesis, compilation, and optimization toolkit that transforms functions from a representation suitable for synthesizing classically-specified circuits to circuit descriptions appropriate for quantum computers \cite{henderson23a,henderson23,sinha23,sinha22,smith19,smith19a}.
There is often more than one way to represent a classically-specified function as a quantum circuit, and previous work has shown that different strategies lend themselves to different circuit properties, including number of required qubits, number of required gates, and required gate types\cite{henderson23a,henderson23,sinha22}.
These features affect the viability of running circuits on contemporary quantum hardware, which currently has relatively few qubits, all of which suffer from significant amounts of noise \cite{preskill18,castelvecchi23,chapman20}.
Therefore, understanding ideal use cases and optimizations for varying quantum circuit types is important, and \textit{MustangQ} automates the rote circuit generation tasks required to efficiently complete such analysis.

Our hash reversal procedure depends upon only a subset of \textit{MustangQ}’s functionality, and for the remainder of this paper, we will discuss only that subset.
These capabilities are motivated by the fact that quantum and classical computers, while both involving circuits, have a number of important differences.
One such difference is that some quantum circuit synthesis methods require reversible function specifications\cite{fazel07}.
As a result, \textit{MustangQ} has the ability to transform irreversible functions into reversible forms. 
So, when hash functions are provided as inputs, \textit{MustangQ} can synthesize reversible function representations that are then embodied in synthesized circuits.

\begin{figure} [ht]
   \begin{center}
   \begin{tabular}{c}
   \includegraphics[width=0.5\linewidth]{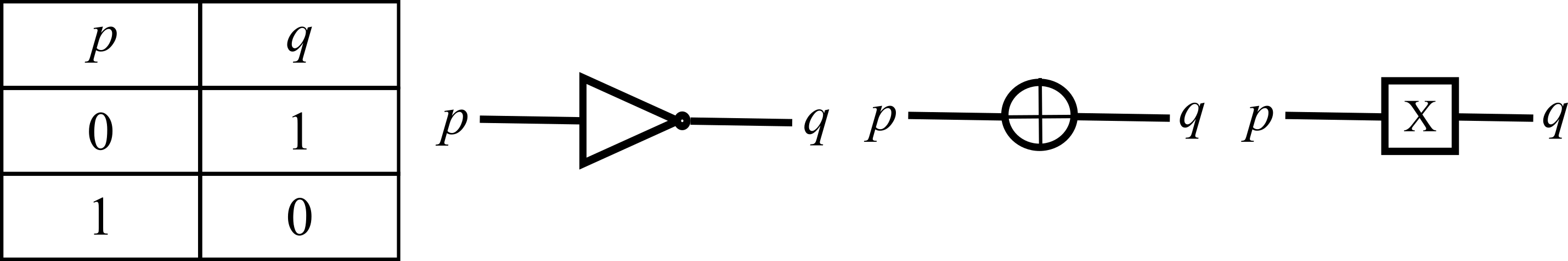}
   \end{tabular}
   \end{center}
   \caption{\label{fig:NOT} 
Truth table and circuit representation for a reversible NOT operation. The first representation is common in classical circuits, while the second and third are used in quantum circuits.}
\end{figure}

\begin{figure} [ht]
   \begin{center}
   \begin{tabular}{c}
   \includegraphics[width=0.35\linewidth]{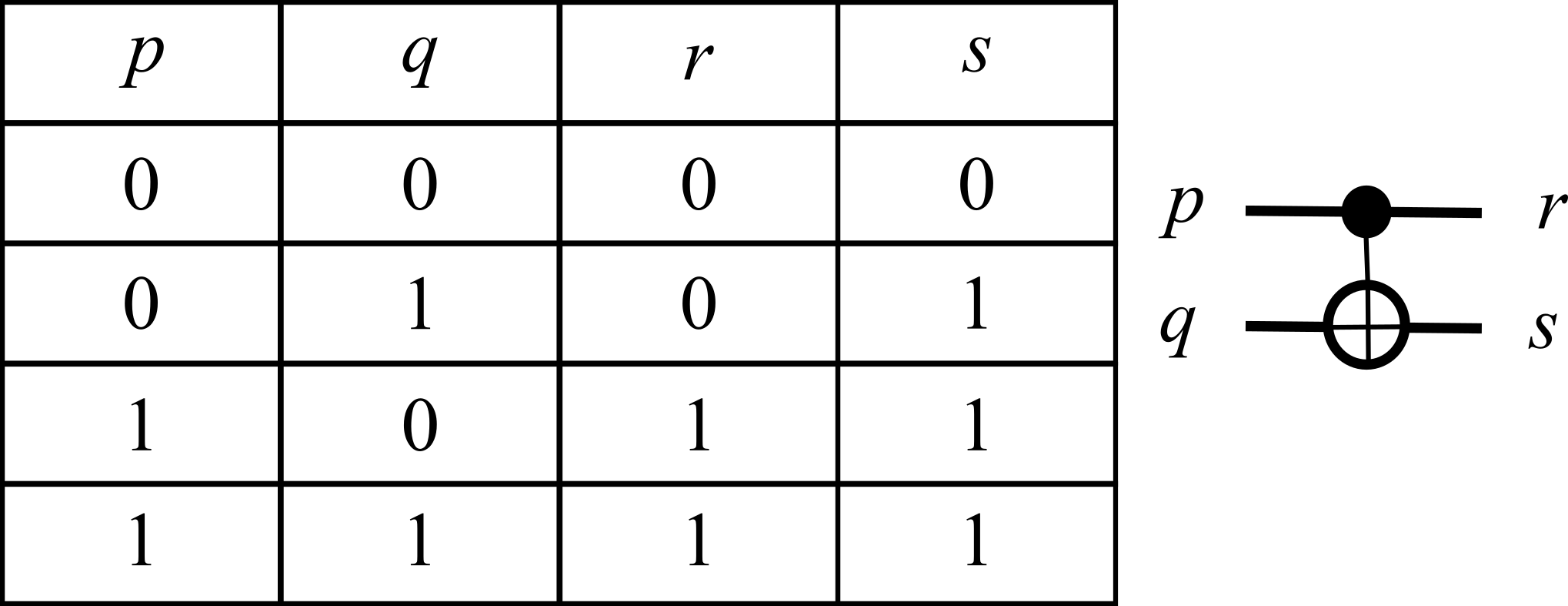}
   \end{tabular}
   \end{center}
   \caption{\label{fig:CNOT} 
Truth table and circuit representation for a reversible Controlled-NOT operation.}
\end{figure}

\begin{figure} [ht]
   \begin{center}
   \begin{tabular}{c}
   \includegraphics[width=0.5\linewidth]{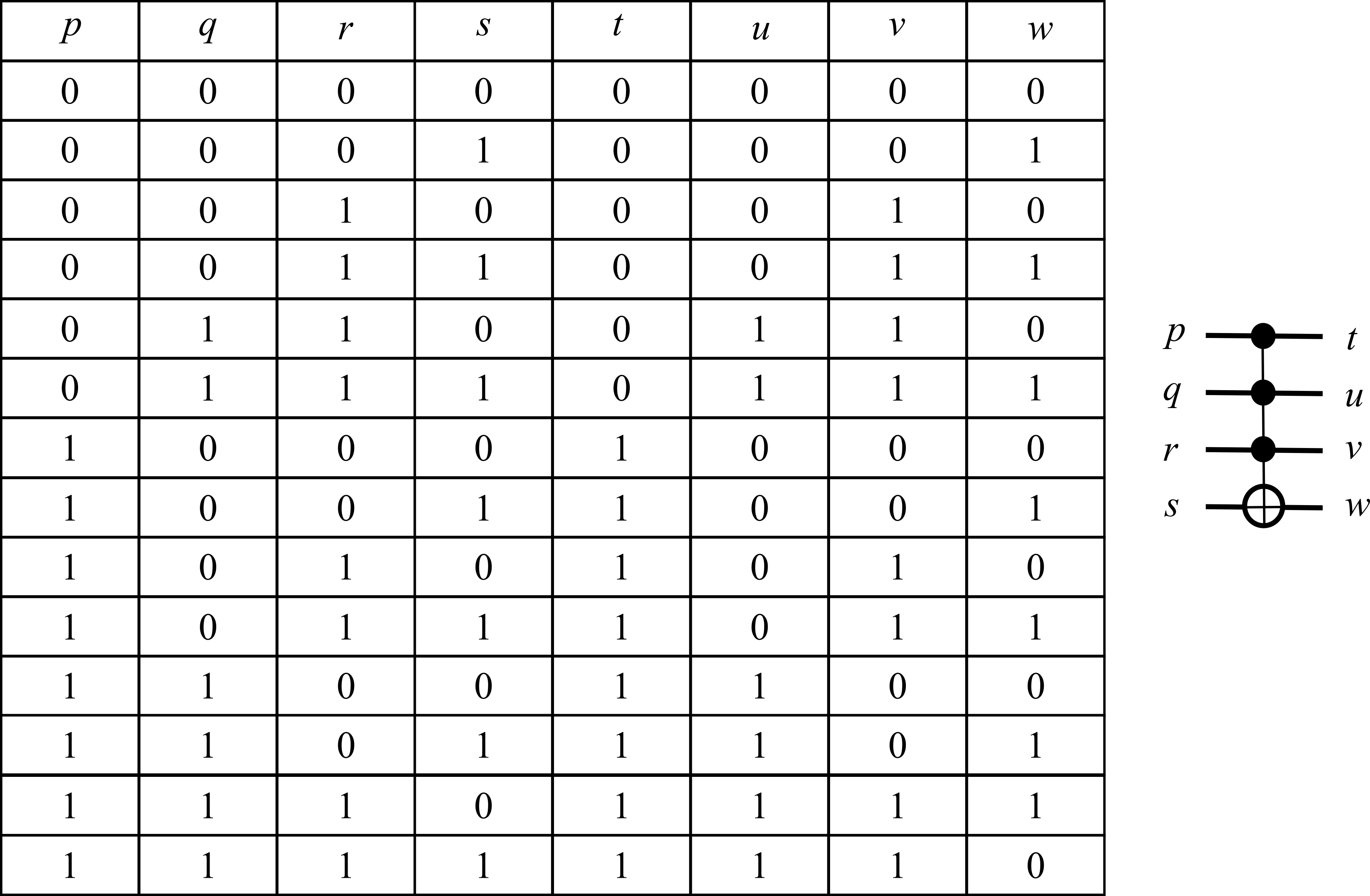}
   \end{tabular}
   \end{center}
   \caption{\label{fig:generalized_Toffoli} 
Truth table and circuit representation for a generalized Toffoli with three controls: $p$, $q$, and $r$ are controls for target $s$; $t$, $u$, and $v$ maintain the values of $p$, $q$, and $r$; and $w = \sim s$ when $p$, $q$, and $r$ are all one.}
\end{figure}

The three gates used in this article for inverse hash circuits are NOT or Pauli-$\mathbf{X}$ gates (illustrated in Figure \ref{fig:NOT}), Controlled-Not or C-NOT gates (illustrated in Figure \ref{fig:CNOT}), and Toffoli gates (illustrated in Figures \ref{fig:Toffoli} and \ref{fig:generalized_Toffoli}).
Toffoli gates are either `true' Toffolis with only two controls (Figure \ref{fig:Toffoli}), or generalized Toffolis with three or more controls (Figure \ref{fig:generalized_Toffoli}).
It is worth noting that this specific gateset is applicable for both quantum and classical circuits.
Thus, while \textit{MustangQ}'s primary use is generation of quantum circuits---meaning this article will use the notation and terminology standard for that field---it is worth noting that the circuits described herein can also be understood classically.
Future work (briefly addressed in Section \ref{sec:conclusion_and_future_work}) ought to investigate how these inverse hash circuits might be used in a quantum-specific way to parallelize the output-to-input procedure.

\section{Computing Inverse Hash Functions}
To obtain a circuit specification that constitutes an inverse hash function, there are three broad steps.
First, prepare the given hash function as an electronic design file.
Second, synthesize a circuit from that function using the ESOP Synthesis Method, which generates reversible circuits even from irreversible function specifications.
Third, reverse the synthesized circuit by simply flipping the sequential order of the gates comprising the circuit.

\subsection{Preparing the Hash Function}\label{sub:preparing_the_function}
The first step is to specify the hash function as an electronic design file.
There are several acceptable formats, including structural Verilog HDL netlists or decision diagrams, such as Binary Decision Diagrams (BDD).
In Section \ref{sec:conclusion_and_future_work}, we briefly discuss such formats, as they present scalable approaches to bitstring function representation.
But in most of this paper, we focus on the \texttt{.pla} format, which was originally devised as a compact textual representation for specifying digital logic programmable logic arrays for physical implementation in electronic circuits and, in particular, as a file format that allowed specifications to be minimized\cite{brayton84,rudell86}.
We selected \texttt{.pla} as the primary format in this article for two reasons.
First, \texttt{.pla} files are relatively straightforward and standardized, providing a user-friendly---and parser-friendly---way to specify bitstring functions.
Second, we have a sizeable set of benchmark function specifications in either \texttt{.pla} format or in other tabular forms that are readily converted to \texttt{.pla} files, including collections maintained via RevLib \cite{wille08}.
So, the first step in computing an inverse hash function is to prepare that function as a \texttt{.pla} file.

The \texttt{.pla} format can describe arbitrary functions of the form $f:\mathbb{B}^n \rightarrow \mathbb{B}^m$, where $\mathbb{B}=\{0,1\}$, and $n$ and $m$ are integers.
Each row of a \texttt{.pla} file consists of $n+m$ bits: the first $n$ bits are one input, and the remaining $m$ bits constitute a corresponding output.
Additionally, each row can represent multiple input/output pairs, using dashes (`-').
Specifically, a dash in an input bitstring indicates that either a $0$ or a $1$ is an input that corresponds to the associated output.
And a dash in an output bitstring indicates that either $0$ or $1$ may be used interchangeably in that particular output.
Thus, functions that have cube-covering overlap are appropriate for efficient representation in \texttt{.pla} format.

\begin{figure} [ht]
   \begin{center}
   \begin{tabular}{c}
   \includegraphics[width=0.5\linewidth]{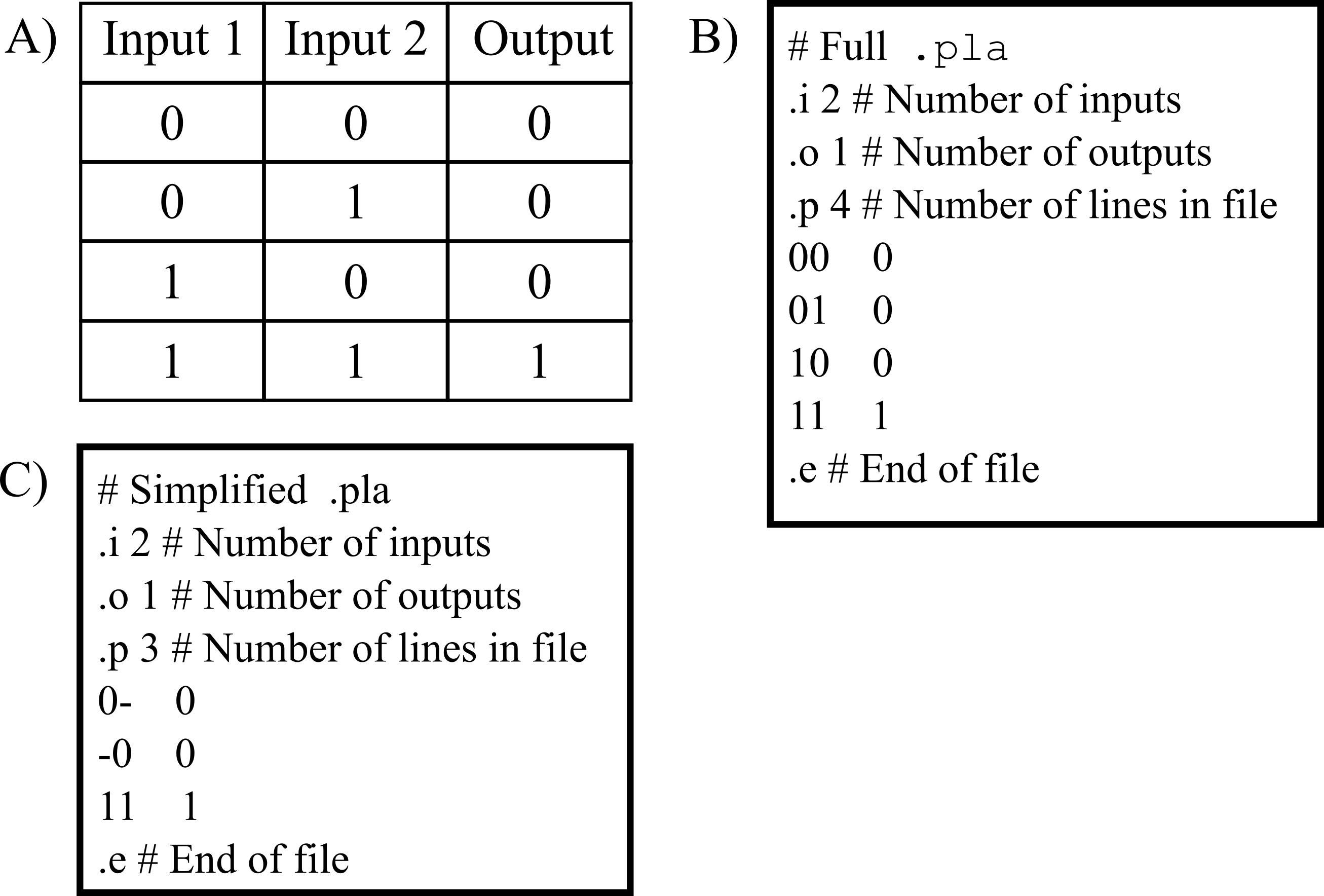}
   \end{tabular}
   \end{center}
   \caption{\label{fig:pla_example} 
Subfigure A illustrates a simple function translated to \texttt{.pla} format in Subfigure B. Subfigure C illustrates a simplification of the \texttt{.pla} with the introduction of hyphens (-). The hyphen in line 1 indicates that, regardless of the value of the second input, when the first input is zero, the output is also a zero. Similarly, the hyphen in line 2 indicates that, whenever the second input is a zero, the output is also zero. For this trivial function, adding such hyphens removes only one line from the \texttt{.pla}, but for larger functions with more overlap, the reduction of the number of lines can be significant.}
\end{figure}

Figure \ref{fig:pla_example} illustrates a trivial function represented as a \texttt{.pla} file.
The full tabular representation in Subfigure B is simplified using dashes to the form in Subfigure C, and while this trivial example has minimal simplification, \texttt{.pla} files that represent larger functions with significant amounts of cube-covering overlap can be significantly reduced in size.
Consequently, using \texttt{.pla} files is advantageous because they are intuitive and can efficiently represent even very large functions, so long as those functions have sufficient cube-covering overlap; for example, \texttt{.pla} files can efficiently represent functions with over one hundred inputs and outputs.\footnote{For example, the file \texttt{frg2} from RevLib contains 143 inputs and 139 outputs\cite{wille08}.}

\subsection{Synthesizing Hash-Function-Representing Circuits}\label{sub:synthesizing_circuits}
After preparing an appropriate function specification, the Exclusive-Or-Sum-of-Products (ESOP) Synthesis Method is used to produce a reversible circuit\cite{fazel07}.
The Method---as with the rest of the \textit{MustangQ} toolkit---was originally developed with quantum circuits in mind, and produces a circuit with the original number of inputs plus the original number of outputs.
The first step is generally to produce a minimized Exclusive-Or-Sum-of-Products, or ESOP, from the \texttt{.pla} representation of a hash function.
An ESOP is a function with inputs and outputs specified as bitstrings whose individual bits are interpreted as being combined via logical conjunctive AND operators.
These bitstrings (termed products) are then summed with disjunctive exclusive-OR operators\cite{fazel07}.
Evaluating the resulting expression with a given input bitstring produces that input's corresponding output bitstring.
Although most functions will not natively be in ESOP form, they can be converted via an automated process, whereupon the rows in a tabular representation of bitstring functions are the products of the ESOP\cite{mishchenko01}.
This process is termed ESOP minimization, and we use EXORCISM-4 to implement it.

The ESOP Synthesis Method produces a quantum circuit by mapping each product to a Toffoli gate in two steps.
First, qubits are added to the circuit: one qubit is added for each input variable, and one qubit is added for each output variable.
Qubits must be added for both the input and output variables, because circuits generated with ESOP synthesis do not overwrite the input values with output values.
Second, for each product in the table, and for each output variable value of 1 therein, a Toffoli gate is added to the circuit.
The controls of the Toffoli gate are determined by the inputs of the current product: an input variable value of 1 means a control is placed on the corresponding input qubit, while an input value of 0 means a negative control is placed on the corresponding input qubit.
A negative control is simply a control with a NOT gate on either side; the NOT gates reverse the qubit’s polarity before the control and restore it after the Toffoli gate has executed.
The Toffoli gate’s target is placed on the output qubit corresponding to the output variable whose value is 1.

Before turning to an example and experimental results, we include two clarifying points.
First, we explain how this process generates reversible circuits.
By using only reversible operations (recall from Section \ref{sub:logical_reversibility} that NOT gates, C-NOT gates, and Toffoli gates are reversible) and a qubit for each input \textit{and} each output, the ESOP Synthesis Method preserves the input values throughout the computation.
This interpretation of the function is consequently reversible, because each set of input values is effectively appended---unchanged---to each set of output values at the end of circuit computation.
Second, we note that the core approach of the ESOP Synthesis Method also works for tabular functions that have not been minimized to ESOP form.
Specifically, the process of adding gates as specified above works to produce reversible circuits from functions in non-minimized \texttt{.pla} form, but at the cost of far more gates.
Consequently, it is generally advantageous to use the ESOP Synthesis Method as intended, with functions in ESOP form; however, minimization is a computationally-intensive process, so it is worth noting that the step could be skipped for large hash functions at the cost of substantially less efficient circuits.

\subsection{Computing the Inverse Hash Function Circuits}
Finally, the circuit specification generated in the above step can simply be reversed, meaning that the gate order is swapped.
For example, if gates are numbered $1,2,3,\dots,n-1,n$, the reversed circuit would have gate order $n,n-1,\dots,3,2,1$.
This reversed circuit then embodies the inverse hash function, as illustrated in the next section.

\section{Example and Experimental Results}
\begin{figure} [ht]
   \begin{center}
   \begin{tabular}{c}
   \includegraphics[width=0.75\linewidth]{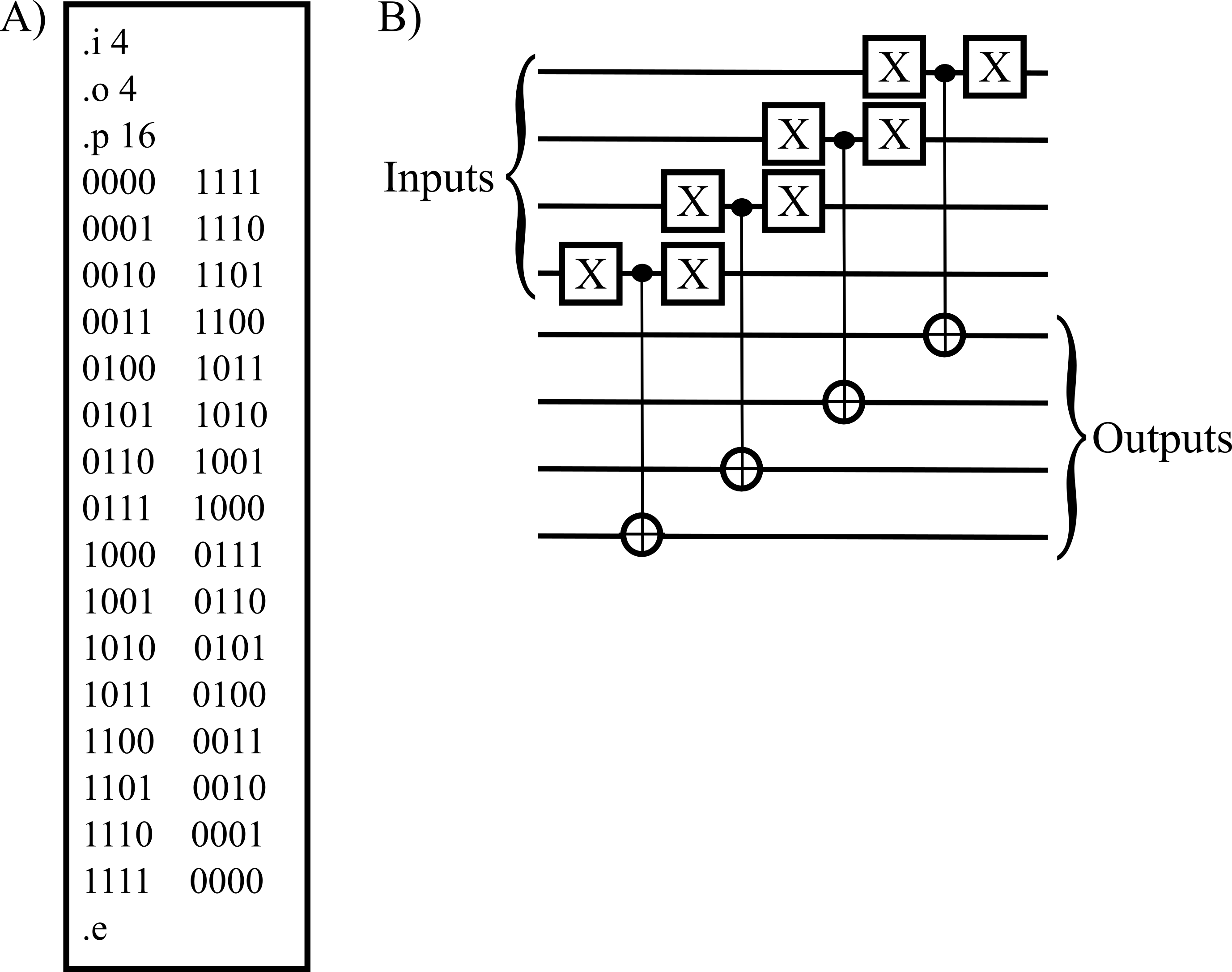}
   \end{tabular}
   \end{center}
   \caption{\label{fig:four_bit_example} 
Subfigure A illustrates a four-bit hash function, while Subfigure B illustrates the circuit synthesized from this function using the ESOP Synthesis Method of \textit{MustangQ}.}
\end{figure}

\begin{figure} [ht]
   \begin{center}
   \begin{tabular}{c}
   \includegraphics[width=0.5\linewidth]{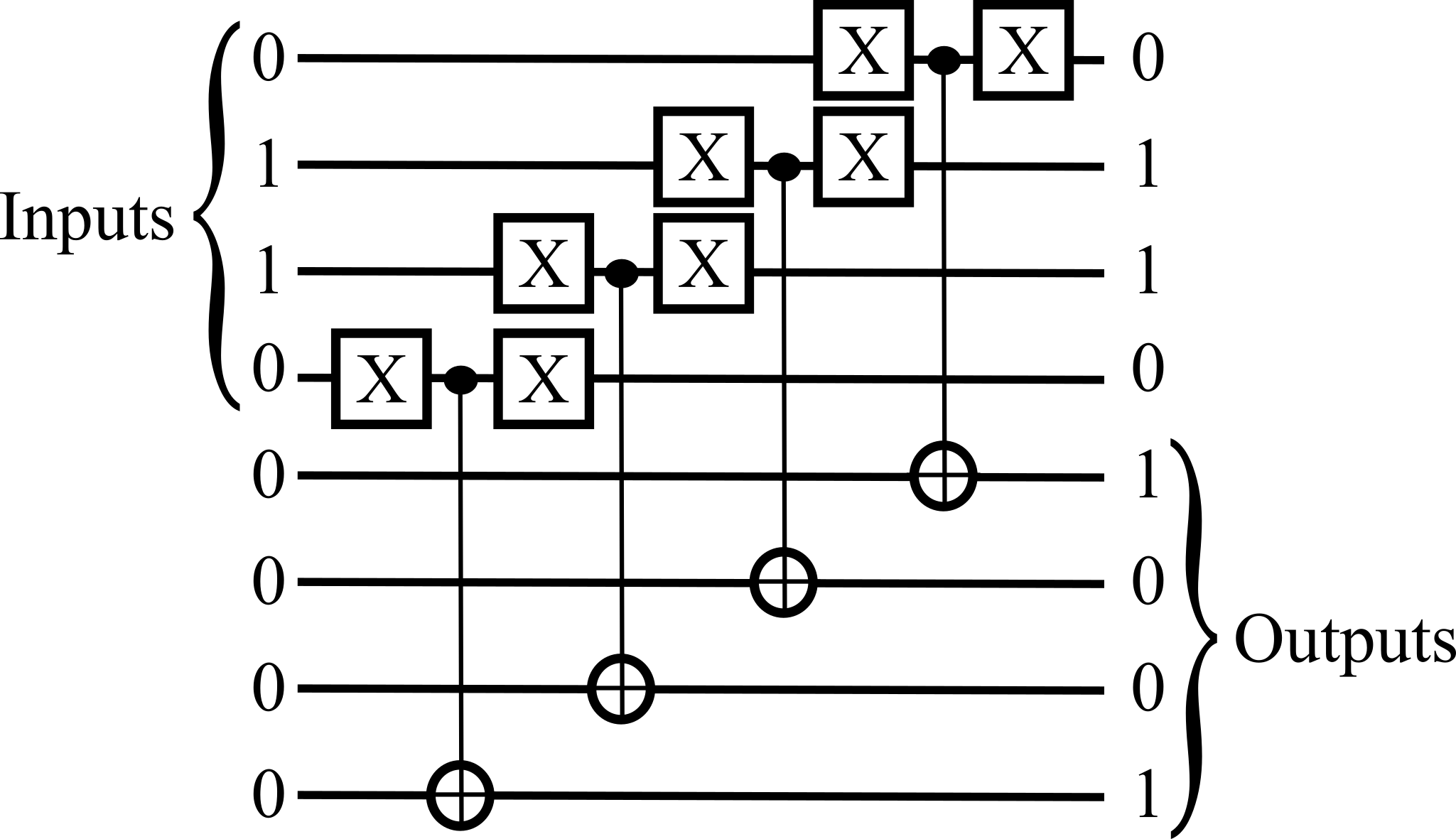}
   \end{tabular}
   \end{center}
   \caption{\label{fig:four_bit_ex_circuit} 
Generating the output 1001 from input 0110. Notice that the inputs remain unchanged, while the outputs contain the desired calculation.}
\end{figure}

\begin{figure} [ht]
   \begin{center}
   \begin{tabular}{c}
   \includegraphics[width=0.5\linewidth]{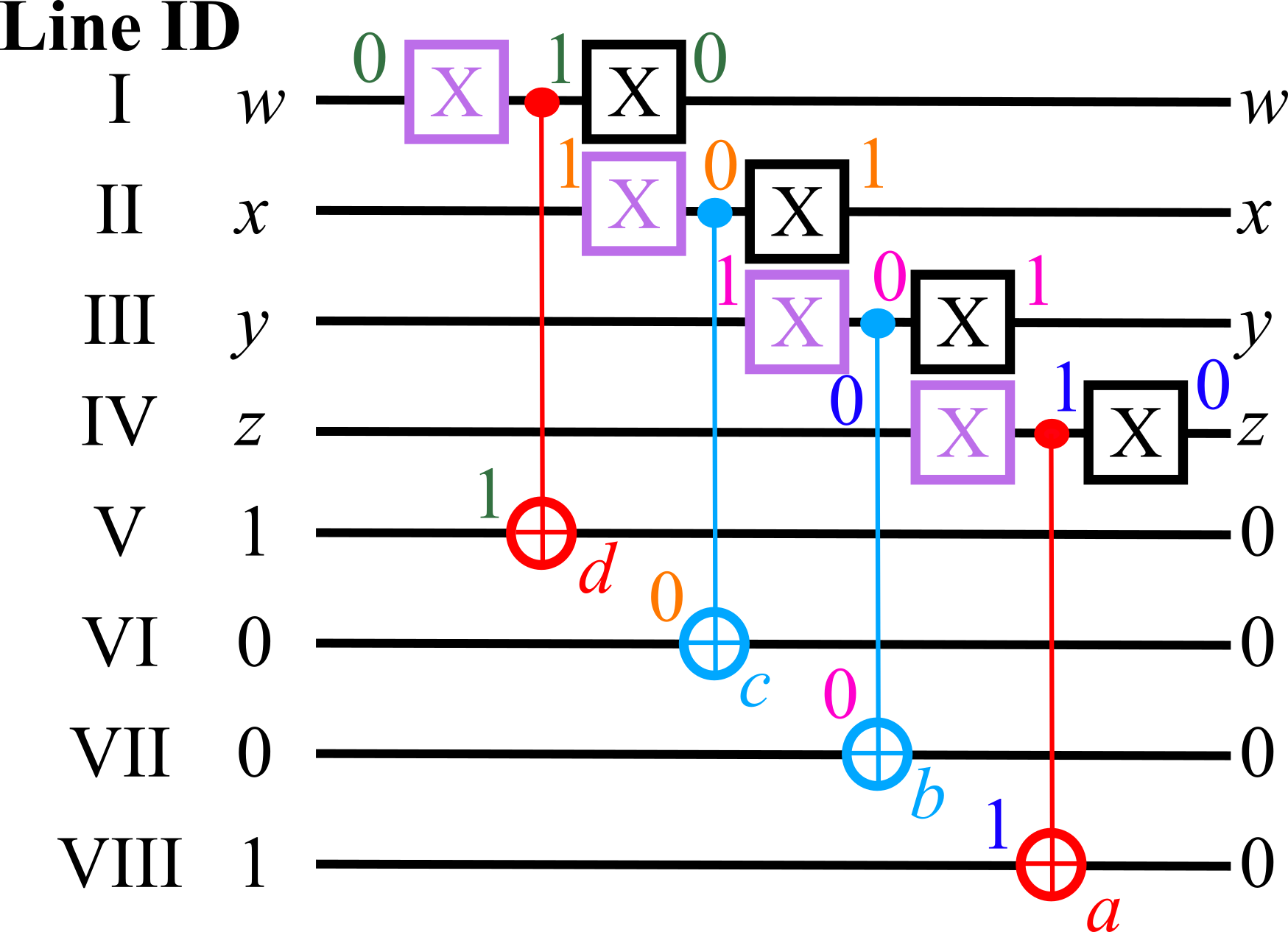}
   \end{tabular}
   \end{center}
   \caption{\label{fig:four_bit_inverse} 
Working backwards from output values 1001 to obtain the input values 0110. The numbers closest to the right-hand side on lines I, II, III, and IV are the deduced input values: $w=0$, $x=1$, $y=1$, and $z=0$. The numbers indicating intermediate values of each line are color-coded such that values related to $w$’s reversal are in pine green, values related to $x$’s reversal are in orange, values related to $y$’s reversal are in pink, and values related to $z$’s reversal are in dark blue. Gates in red denote that the gate flipped its output value, while gates in light blue denote that the output was not changed. The accompanying red and light blue letters serve to identify particular gates in the text description.}
\end{figure}

We begin this section with an example of synthesizing a circuit for a trivial four-bit hash function.
Subfigure A of Figure \ref{fig:four_bit_example} presents the hash function in tabular form; it is an extremely weak hash function, both because of its size and because it does not satisfy the Avalanche Criteria.
But for purposes of considering an illustrative example, we did not need to craft a strong hash function because---besides requiring a longer completion time and larger circuits---the process does not change with hash function strength.
Applying \textit{MustangQ}’s ESOP Synthesis Method produces the circuit of Subfigure B of Figure \ref{fig:four_bit_example}.
Figure \ref{fig:four_bit_ex_circuit} again illustrates the circuit, this time with the input and output values labelled with a particular input/output pair: $0110$ and $1001$.
As noted in Section \ref{sub:synthesizing_circuits}, the input values pass through the circuit unchanged, while the outputs evolve from a state of all zeros to the values associated with the given input.
Reversing the order of the gates in the circuit specification allows for working backwards from the output values to obtain the input values.
This input-retrieval process is illustrated in Figure \ref{fig:four_bit_inverse}, to which we now turn.

We begin on the left-hand side, with variables $w$, $x$, $y$, and $z$ representing the unknown input values.
The numerical values on the left-hand side represent the known output values corresponding to the input values we seek.  On the right-hand side, variables $w$, $x$, $y$, and $z$ again represent those unknown input values, while the initial output values are set to zero.

It is worth emphasizing two points.
First, the values of $w$, $x$, $y$, and $z$ are the same on both the left and right sides of the circuit, because the paired NOT gates restore input values after changing them.\footnote{The NOT gates that occur to the left of the C-NOT gates in Figure \ref{fig:four_bit_inverse} are the input restoration gates that are not strictly necessary; in Figure \ref{fig:four_bit_inverse}, these gates are lavender-colored.}
While this input immutability is not a requirement for the reversal process, it is a feature of the \textit{MustangQ} toolkit that is useful in synthesizing circuits more generally, so we leverage it here.
Second, the initialization of output values to zero is not a requirement for function reversal, but is again a \textit{MustangQ} feature that is also convenient here.
It would be possible to use this same reversal procedure with output values initialized to non-zero values; what is important is that the initialization values be a known part of the synthesis process.
The input values can be determined by working backwards; because the circuits produced from reversible functions are reversible, we can ascertain what must have happened at each step to obtain the known outputs.
Figure \ref{fig:four_bit_inverse} denotes the numbers that are relevant to obtain each of the four input values; for the sake of brevity, we describe the process for only two of the four.

Consider the red gate identified with a red \textit{a}.
Line VIII has a final output value of 1, but it started as a 0.
By definition of the C-NOT operation, this means that red gate \textit{a} flipped the value of line VIII, such that the control value must have been a 1, denoted by the dark blue 1 to the right of the control on line IV.
Tracing line IV to the right, we find a NOT gate.  Since the result of that gate must be a 1, we know that the input to the NOT was 0, and consequently, we find that input $z$ had a starting value of 0.

As our second example, consider the light blue gate identified as \textit{b}.  Blue gate \textit{b} has a target on line VII which has both a final and starting value of 0.
This means that the control for blue gate \textit{b} (located on line III), must have had an input of 0, denoted by the pink zero to the right of the control.
Tracing from that control to the beginning of line III leads to a NOT gate; since the result of the NOT is 0, the input must have been 1, meaning that $y=1$.

Such manual validation is possible only for very small functions.
For even moderately larger functions (such as those in Table \ref{tab:results}, validation was accomplished by combining the circuit specification in the ‘forward’ (\textit{i.e.}, input to output) direction with that for the ‘reversed’ (\textit{i.e.}, output to input) direction.
We used Verilog HDL simulation to verify that all of the values entered on the left side matched those on the right side; when the combined forward and reversed circuits act as identity for all possible input/output pairs, reversal is proven successful.
We have inverted the hashes presented in Table \ref{tab:results} and have validated the results for each.

The entries in Table \ref{tab:results} are all very small hash functions that are intended as preliminary, proof-of-concept results; they are sufficiently small that we were able to use the \texttt{.pla} format, to validate exhaustively, and even to manipulate a few of the circuits by hand during development.
These results illustrate that our approach is worthy of further exploration for four reasons.

First, note that none of these times are prohibitive, nor even close: the average ESOP minimization time is about 0.46 seconds, the average circuit synthesis time after minimization is about 0.089 seconds, and the average circuit synthesis time without minimization is about 0.256 seconds.
All of the known temporal results in the table require under ten minutes of CPU time, and this is with a preliminary implementation that could undoubtedly be optimized, including via more efficient data structures.
Such optimization is important, as it is worth noting that the wall clock times for processing these circuits are often quite a bit longer than the required CPU time; for example, the 8-bit hash that was designed to satisfy the Avalanche Criteria, the wall clock time is on the order of minutes.
This is a consequence of time-consuming file I/O operations that are in part due to the human-readable format of \texttt{.pla} files, as well as the frequent write operations used to validate the circuit synthesis and report on its statistics.
Consequently, we are motivated to explore alternative formats for representing and processing the functions, as described in Section \ref{sec:conclusion_and_future_work}.

Second, it is worth noting that although these hash functions are small, \textit{MustangQ} has been successfully tested at scale in situations where the functions are not hash functions\cite{henderson23a,henderson23,sinha22}, and we have yet to test a function for which \textit{MustangQ} cannot generate a circuit.
While this is certainly not to say that such functions do not exist, it is worth emphasizing that \textit{MustangQ}'s capabilities are proven far beyond functions with 8 inputs and outputs\cite{henderson23a,henderson23,sinha22}.
Additionally, even with hash functions, we have tested \textit{MustangQ} with slightly larger `hashes' of up to 21 bits.
So, although we opted not to report those results here (the hashes are trivial, with outputs that are simply offsets of the inputs), we emphasize that the experiments of Table \ref{tab:results} do not express the extent of \textit{MustangQ}'s capabilities.
What does remain to be shown is that our method for inverting hashes will scale as larger hash functions---and not just larger functions in general---are synthesized via \textit{MustangQ}.

Third and related to minimization, it is noteworthy that adding bits---and complexity of the function---causes a steep increase in the amount of time required for ESOP minimization.
Although the total number of inputs and outputs increases by a factor of at most 2, the corresponding increase in ESOP minimization time is a factor of at least 172.
This increase seems more substantial when contrasted with the time increase required for the circuit synthesis process itself.
Specifically, for construction of the circuit after ESOP minimization, there is a factor of about 58 between the largest and smallest times, and for circuit synthesis without ESOP minimization, there is a factor of about 61 between the same.
As both of these are an order of magnitude below the corresponding increase for ESOP minimization time, optimizing or eliminating the ESOP minimization may be necessary when scaling the inverse hash computation procedure.

Fourth and finally, we note that although ESOP minimization does impose a temporal cost, it offers meaningful gains in the number of gates required.
Synthesizing without first minimizing always imposed a gate count penalty, and the average number of gates per circuit increased from about 273 with minimization to about 514 without.

\begin{table}[ht]
\caption{Statistics for hash function circuits with five notes: 1) The 8-bit hash was custom designed to satisfy the Avalanche Criteria. 2) The time to synthesis was recorded as CPU time when running on a HP DL380 server with a 16 Core Intel Xeon processor with 2.9GHz and 380GB of RAM. 3) The first set of times and gate counts is from runs using ESOP minimization.  As such minimization could be removed if scalability were an issue, the second set of times and gate counts are for synthesis without ESOP minimization. 4) All synthesis results include a standard optimization that removes superfluous NOT gates. 5) EXORCISM-4 reports timing results in intervals of 0.01 seconds. So, ``$<$ 0.01 s'' indicates when the minimization was faster than that time, and the ``*'' indicates that EXORCISM-4 ran successfully, but reported its time-counter had overflowed.} 
\label{tab:results}
\begin{center}
\begin{tabularx}{\textwidth} { 
  | >{\centering\arraybackslash}X 
  | >{\centering\arraybackslash}X 
  | >{\centering\arraybackslash}X
  | >{\centering\arraybackslash}X
  | >{\centering\arraybackslash}X
  | >{\centering\arraybackslash}X
  | >{\centering\arraybackslash}X 
  | >{\centering\arraybackslash}X |}
\hline
\rule[-1ex]{0pt}{3.5ex}  Function Name & Number of Inputs & Number of Outputs & ESOP Minimization Time & Circuit Synthesis Time (With Minimization) & Circuit Gate Count (With Minimization) & Circuit Synthesis Time (No Minimization) & Circuit Gate Count (No Minimization)  
\\
\hline
\rule[-1ex]{0pt}{3.5ex}  4-bit AES S-Box & 4 & 4 & $<$ 0.01 s & 0.005888 s & 29 & 0.015233 s & 60 \\
\hline
\rule[-1ex]{0pt}{3.5ex}PRESENT S-Box & 4 & 4 & $<$ 0.01 s & 0.006541 s & 29 & 0.015096 s & 60 \\
\hline
\rule[-1ex]{0pt}{3.5ex}  DES S-Box 1 & 6 & 4 & 0.01 s & 0.025588 s & 105 & 0.071742 s & 246 \\
\hline
\rule[-1ex]{0pt}{3.5ex}  DES S-Box 2 & 6 & 4 & 0.01 s & 0.019397 s & 86 & 0.072220 s & 246 \\
\hline
\rule[-1ex]{0pt}{3.5ex}  DES S-Box 3 & 6 & 4 & $<$ 0.01 s & 0.017789 s & 82 & 0.071665 s & 246 \\
\hline
\rule[-1ex]{0pt}{3.5ex}  DES S-Box 4 & 6 & 4 & 0.01 s & 0.025794 s & 101 & 0.072256 s & 249 \\
\hline
\rule[-1ex]{0pt}{3.5ex}  DES S-Box 5 & 6 & 4 & 0.01 s & 0.025501 s & 113 & 0.071608 s & 246 \\
\hline
\rule[-1ex]{0pt}{3.5ex}  DES S-Box 6 & 6 & 4 & 0.01 s & 0.023222 s & 90 & 0.071501 s & 246 \\
\hline
\rule[-1ex]{0pt}{3.5ex}  DES S-Box 7 & 6 & 4 & $<$ 0.01 s & 0.023332 s & 90 & 0.071921 s & 246 \\
\hline
\rule[-1ex]{0pt}{3.5ex}  DES S-Box 8 & 6 & 4 & 0.01 s & 0.016621 s & 103 & 0.071770 s & 248 \\
\hline 
\rule[-1ex]{0pt}{3.5ex}  AES S-Box & 8 & 8 & 1.72 s & 0.326583 s & 935 & 0.916981 s & 1532 \\
\hline
\rule[-1ex]{0pt}{3.5ex}  AES Inverse S-Box & 8 & 8 & 1.29 s & 0.301835 s & 869 & 0.882464 s & 1532 \\
\hline
\rule[-1ex]{0pt}{3.5ex}  8-bit hash & 8 & 8 & 1.06 s & 0.344246 s & 918 & 0.925617 s & 1532 \\
\hline
\end{tabularx}
\end{center}
\end{table}

\section{Conclusion and Future Work}\label{sec:conclusion_and_future_work}
Efficiently reversing hashes has a number of applications, and while some lend themselves to more pernicious consequences than others, all of them are important to the cybersecurity community.
Consequently, this article is an important step that also raises several avenues of further research, which are primarily focused on the goal of testing the procedure at scale.
We close by briefly considering two: further exploration of applications---including how the relevant hashes might be represented in a tabular format for \textit{MustangQ}---and further exploration of scalable function representation methods.

\subsection{Exploring Applications}
As is widely recognized in works such as Refs. \citenum{preneel10,paar10,mironov05,sobti12,zheng18,chaudhary19,wohlmacher00} and their hundreds of references, hash functions have a host of applications, from participating in blockchain ecosystems (including cryptocurrency mining) to decrypting website credentials to impersonating cryptographic signatures.
\textit{MustangQ}'s circuit generation directly calculates an inverse hash, thus bypassing the massive set of forward-hash computations involved in a brute-force search for any of these applications.
But leveraging the toolkit's full value may require situation-dependence.
For example, the \textit{MustangQ} approach is especially suitable for tasks involving fixed-size hashes---such as those used in blockchains---because the hash reversal circuit would not need to be resynthesized in real time.

To illustrate this, we briefly describe how \textit{MustangQ} might apply to a blockchain application, thus motivating the need for further application-specific research.
Consider cryptocurrency mining, which involves attempting to reverse a hash function that is calculated on a specific blockchain database.
Once the hash function has been reversed, it is used to create a ``Proof-of-Work" (PoW) that validates the integrity of the given blockchain entry.
This mining process typically requires significant computational resources, given that classical computers are often required to perform a brute-force search---which requires computing all of the possible forward hash results for a given blockchain entry---to obtain the reverse of the blockchain signature.
As mentioned above, applying our method would avoid such an exhaustive search.
Furthermore, if the inverse hash circuit were utilized as a quantum---and not a classical---circuit, it might be possible to obtain information about multiple sets of inputs simultaneously, and the feasibility of this on an application-specific basis could prove fruitful.

\subsection{Exploring Alternative Function Representations}
There are several reasons that alternative function representations are important for our procedure, including the spatial feasibility of representing the functions, the temporal feasibility of parsing and pre-processing the functions, and the temporal feasibility of then applying the actual circuit synthesis algorithm.
We have not yet encountered issues with the algorithmic approaches of \textit{MustangQ}\cite{henderson23a,henderson23,sinha23,sinha22,smith19,smith19a}, so here we will focus on the first two issues.

First, space constraints: there are many functions that are too large to be efficiently processed as \texttt{.pla} files.
Despite the size reduction available to functions with cube-covering overlap, \texttt{.pla} files are inherently limited, because the size of \texttt{.pla} specifications for functions without substantial cube-covering overlap grows exponentially with respect to the number of inputs and outputs.
As hash functions may not have substantial overlap in input/output pairs, only hashes with small numbers of inputs and outputs may be represented as \texttt{.pla} files.\footnote{For example, a hash function with an input/output size of 256 bits could require $1.16e77$ lines in a \texttt{.pla}. Quite clearly, that is impractical.}

Therefore, another avenue for future work is to explore efficient production of hash functions represented as either Verilog HDL netlists or decision diagrams, and to add circuit synthesis functionality from such representations to \textit{MustangQ}.
Structural Verilog HDL netlists allow for representing much larger functions than \texttt{.pla} files can accommodate, because they allow for multi-level specifications that do not require explicitly listing each input/output pair, or even a significant fraction thereof in most cases\cite{reese22}. 
Similarly, several forms of decision diagram, including BDDs\cite{bryant86,wille08}, can represent bitstring functions in such a way that the function can be efficiently divided into discrete portions.
If each of these portions could be efficiently synthesized into reversible circuits and put back together appropriately, then decision diagram representations might allow for inverting even very large hash functions.

To date, we have reversed only small hash functions represented as \texttt{.pla} files, because we do not have access to hash functions specified as Verilog HDL netlists of combinational gates nor as decision diagrams.
As obtaining such representations is not trivial, future work ought to develop such representations up to and ideally including a 256-bit hash.

Second, temporal parsing and processing constraints.
The \texttt{.pla} format has served us very well in development of \textit{MustangQ}, as it is human-readable, which not only allows for straightforward development and testing, but also for clear illustration of the processes and optimizations that the toolkit is capable of.
Furthermore, it has been incredibly important to have ready functions for testing that are provided in a standardized---or at least standardizable---format.
Consequently, we have no plans to abandon \texttt{.pla} representation.

However, as has been mentioned several times throughout this article, \texttt{.pla} functions have limited spatial efficiency and, in part because of their human-readability, are not efficient to parse or write.
Currently, our procedure is implemented with the need to read and write \texttt{.pla} files at several points, including to interface with the ESOP minimization step.
Alternative data structures---including decision diagrams---should allow for a more direct interfacing that would require fewer computationally-expensive read/write operations.
So, although \textit{MustangQ} currently does leverage an internal data structure for representing the content of \texttt{.pla} files for as much processing as possible, exploring additional data structures should allow for faster processing times that therefore give a more practical sense of the scalability of our procedure.

\appendix

\bibliography{references} 
\bibliographystyle{spiebib}

\end{document}